\documentclass[twocolumn,prl,showpacs,aps,superscriptaddress]{revtex4-1}

\usepackage[english]{babel}
\usepackage[ansinew]{inputenc}
\usepackage{times}
\usepackage{graphicx}
\usepackage{graphics}
\usepackage{amsmath}
\usepackage{amsfonts}
\usepackage{amssymb}
\usepackage{amsbsy}
\usepackage{dsfont}
\usepackage{epstopdf}
\usepackage{makeidx}
\usepackage{subfigure}
\usepackage{color} 
\usepackage{pgf}
\usepackage{bm}
\usepackage{tikz} \usepackage[normalem]{ulem}

\newcommand{\be}{\begin{equation}}
\newcommand{\ee}{\end{equation}}
\newcommand{\bea}{\begin{eqnarray}}
\newcommand{\eea}{\end{eqnarray}}

\makeindex
\begin{document}

\title{Quantized beam shifts}

\author{W. J. M. Kort-Kamp}
\affiliation{Center for Nonlinear Studies, MS B258, Los Alamos
National Laboratory, Los Alamos, New Mexico 87545, USA}
\affiliation{Theoretical Division, MS B213, Los Alamos
National Laboratory, Los Alamos, New Mexico 87545, USA}
\author{N. A. Sinitsyn}
\affiliation{Theoretical Division, MS B213, Los Alamos
National Laboratory, Los Alamos, New Mexico 87545, USA}
\author{D. A. R. Dalvit}
\affiliation{Theoretical Division, MS B213, Los Alamos
National Laboratory, Los Alamos, New Mexico 87545, USA}
\date{\today}

\begin{abstract}
We predict quantized Imbert-Fedorov, Goos-H\"anchen, and  photonic spin Hall shifts for light beams impinging on a graphene-on-substrate system in an external magnetic field. 
In the quantum Hall regime the Imbert-Fedorov and photonic spin Hall shifts are quantized in integer multiples of the fine structure constant $\alpha$, while the Goos-H\"anchen ones in multiples of $\alpha^2$. 
We investigate the influence on these shifts of magnetic field, temperature, and material dispersion and dissipation. An experimental demonstration of quantized beam shifts could be achieved at terahertz frequencies for moderate values of the magnetic field. 
\end{abstract}
%
\maketitle
Reflection and refraction of light are among the most common phenomena in optics. For a plane wave impinging on an interface separating two media, the propagation of the reflected and transmitted waves is governed by the Fresnel and Snell laws~\cite{BornWolf}. However, this standard geometric optics picture does not apply for a beam of finite width consisting of the superposition of several plane wave components. In this case spatial and angular deviations from the expected ray trajectories occur, resulting in beam shifts within and transverse to the incidence plane, respectively called Goos-H\"anchen (GH) \cite{Goos1947}  and Imbert-Fedorov (IF) \cite{Fedorov1955, Imbert1972} shifts. Even though the spatial IF shift vanishes for transverse electric or transverse magnetic linearly polarized light, photons with opposite heliticities are still  shifted to distinct edges of the reflected/transmitted beam cross section -  the spin Hall effect of light (SHEL) ~\cite{Onoda2004, Bliokh2006, Bliokh2007,Hosten2008, Zhuo2013}. These shifts are relevant for biosensing \cite{Yin1989} and nano-probing \cite{Herrera2010}, and
have been studied for a variety of beam profiles and material media \cite{Pfleghaar1993, Emile1995, Felbacq2004, Schomerus2006, Aiello2008, Aiello2009,  Aiello2010, Haan2010,  Aiello2012, Bliokh2013,  Bliokh2013}. In particular, the influence of gated graphene on beam shifts has been recently investigated \cite{Zhou2012,Grosche2015,Hermosa2015}, and a giant spatial GH shift has been measured \cite{Li2014}.

Here, we show that the magneto-optical response of a graphene-on-substrate system in the presence of an external magnetic field strongly affects beam shifts. In the quantum Hall regime characterized by well-resolved Landau levels in graphene, the IF and SHEL shifts are quantized in integer multiple of the fine structure constant $\alpha=e^2/4 \pi\varepsilon_0 \hbar c$, while the GH shifts are quantized in integer multiples of $\alpha^2$. Disorder broadening of inter-level transitions results in the IF, GH, and SHEL shifts to exhibit a discontinuous behavior at moderate magnetic fields reflecting the discrete Landau-level filling factor. Furthermore, due to time-reversal symmetry breaking, for linearly-polarized incident light the IF shifts change sign when the direction of the applied magnetic field is reversed, while the other shifts remain unchanged. Finally, we discuss the effects of temperature, dispersion, and the role of the \linebreak substrate in this problem.

Let us consider a monochromatic (frequency $\omega$) Gaussian wave-packet propagating in air and impinging at an angle $\theta$ on a non-magnetic, isotropic, and homogeneous substrate of permittivity $\varepsilon$. A graphene sheet is placed on top of the substrate, and a static and uniform magnetic field ${\bf B}$ is applied orthogonal to the graphene-substrate interface (Fig. 1). We assume that the incident beam is confined perpendicularly to the incidence plane, which allows us to neglect the GH shifts and consider only the IF and SHEL ones (we will separately treat the GH shifts at the end of the paper). The incident electric field is given by
${\bf E}_{i}= A(y_i,z_i) [f_p \hat{{\bf x}}_i + f_s \hat{{\bf y}}_i - i f_s  k_0 y_i (\Lambda+ik_0 z_i)^{-1} \hat{{\bf z}}_i] $, 
where $A(y,z)= [2/\pi w_0^2 (1+k_0^2 z^2/\Lambda^2)]^{1/4} e^{i k_0 z - k_0^2y^2/2(\Lambda+ik_0 z)}$ is the Gaussian amplitude, $k_0=\omega/c$ is the magnitude of the wave-vector, and $\Lambda=k_0^2w_0^2/2$ is the dimensionless Rayleigh range of the beam with waist $w_0$ \cite{BornWolf}. The polarization of the incident beam is given by ${\bf \hat{f}}= f_{p}{\bf\hat{x}}_{i}+f_{s}{\bf\hat{y}}_{i}$, where $f_{p}$ and $f_{s}$ are complex amplitudes ($|f_{p}|^2 + |f_{s}|^2 = 1$).
Unit vectors $({\bf\hat{x}}_i, {\bf\hat{y}}_i,{\bf\hat{z}}_i)$ are associated to a reference frame $(x_i, y_i, z_i)$ attached to the central component of the incident beam with origin at the point where the latter reaches the surface. Employing standard Fresnel reflection matrices 
for each component of the incident beam \cite{Tse2011}, one can compute the reflected beam in the paraxial approximation following \cite{Bliokh2006,Xu2011}:
\begin{eqnarray}
{\bf E}_r &=&  A(y_r, z_r)  (f_p r_{\!_{pp}} + f_s  r_{\!_{ps}} ) \Bigg\{\left[1-\dfrac{i C k_0 y_r  Y_p r_{\!_{pp}} } {f_p r_{\!_{pp}} + f_s r_{\!_{ps}} } \right] \hat{{\bf x}}_r \nonumber \\
&& + m \left[1-\dfrac{i C k_0 y_r Y_s r_{\!_{ss}} } {f_s r_{\!_{ss}} + f_p r_{\!_{sp}}} \right]\hat{{\bf y}}_r - m C y_r  \hat{{\bf z}}_r\Bigg\} .
\label{Field}
\end{eqnarray}
The reference frame $(x_r, y_r, z_r)$ has the same origin as 
$(x_i, y_i, z_i)$ with unit vectors  ${\bf\hat{x}}_r={\bf\hat{x}}_i - 2 {\bf\hat{x}}_L ({\bf\hat{x}}_i \cdot {\bf\hat{x}}_L)$, 
${\bf\hat{y}}_r={\bf\hat{y}}_i$, and
${\bf\hat{z}}_r={\bf\hat{z}}_i - 2 {\bf\hat{z}}_L ({\bf\hat{z}}_i \cdot {\bf\hat{z}}_L)$ [${\bf\hat{x}}_L$ and ${\bf\hat{z}}_L$ are lab frame versors, see Fig. 1].
In Eq.(\ref{Field}), $r_{\!_{ij}}$ are the graphene-on-substrate reflection amplitudes for incoming $j-$ and outgoing $i$-polarization $(i,j=s,p)$,
$m = (f_s r_{\!_{ss}} + f_p r_{\!_{sp}}) / (f_p r_{\!_{pp}} + f_s r_{\!_{ps}})$, $C = ik_0/(\Lambda+ik_0z_r)$, 
$Y_p = i \cot(\theta) f_s (r_{\!_{pp}}+ r_{\!_{ss}})/ k_0 r_{\!_{pp}}$, and $Y_s = -Y_p|_{p \leftrightarrow s}$.
The magneto-optical response of graphene in the presence of the magnetic field results in polarization conversion and strongly
affects the profile of the reflected beam.
\begin{figure}
\centering
\includegraphics[scale=0.45]{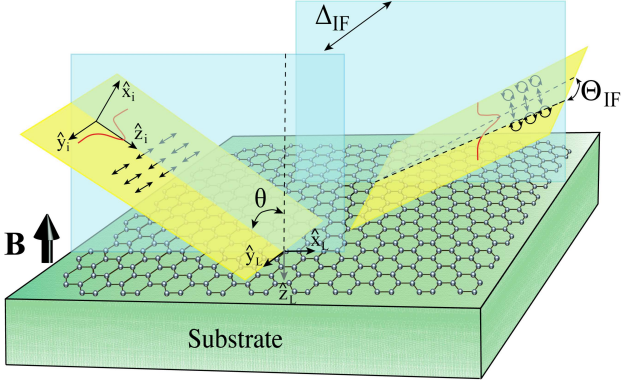}
\caption{Imbert-Fedorov and spin Hall effect of light in a graphene-substrate system in the presence of an external magnetic field.}
\label{Fig1}
\end{figure}

The shifts of the reflected beam can be obtained by calculating the intensity distribution centroid in the $(x_r, y_r, z_r)$ frame \cite{Bliokh2013,Aiello2012}.  
The spatial and angular IF shifts are respectively given by $\Delta_{\textrm{IF}} = \textrm{Re}[\tilde{\Delta}_{\textrm{IF}}]$ and 
$\Theta_{\textrm{IF}} =(k_0/\Lambda) \textrm{Im}[\tilde{\Delta}_{\textrm{IF}}]$, where
\begin{equation}
\tilde{\Delta}_{\textrm{IF}} = \dfrac{Y_p  r_{\!_{pp}}(f_{p}^{*}r_{\!_{pp}}^{*}+f_{s}^{*}r_{\!_{ps}}^{*})}{|f_{s}r_{\!_{ss}} + f_{p}r_{\!_{sp}}|^2+ |f_{p}r_{\!_{pp}}+f_{s}r_{\!_{ps}}|^2} + \{ p\leftrightarrow s \}\, .
\label{ComplexDelta}
\end{equation}
In contrast to isotropic materials, in which $\Delta_{\textrm{IF}}=\Theta_{\textrm{IF}} = 0$ for $s$- or $p$-linearly polarized light~\cite{Bliokh2013}, the magnetic field induces anisotropic optical response in graphene and 
allows for non-vanishing IF shifts even for purely $s$ or $p$ polarizations. Indeed, setting $f_s=1$ or $f_p=1$ in Eq. (\ref{ComplexDelta}) one sees that 
$\tilde{\Delta}_{\textrm{IF}}$ becomes proportional to the cross-polarization reflection coefficient $r_{\!_{ps}}$ which, in turn, is proportional to the Hall conductivity of graphene $\sigma_{xy}$ \cite{Tse2011}.
Since the latter is an odd function of $B$ (due to time-reversal symmetry breaking), it follows that the IF shifts change sign under inversion of the magnetic field.

\begin{figure}
\centering
\hspace{-0.2cm}\includegraphics[scale=0.45]{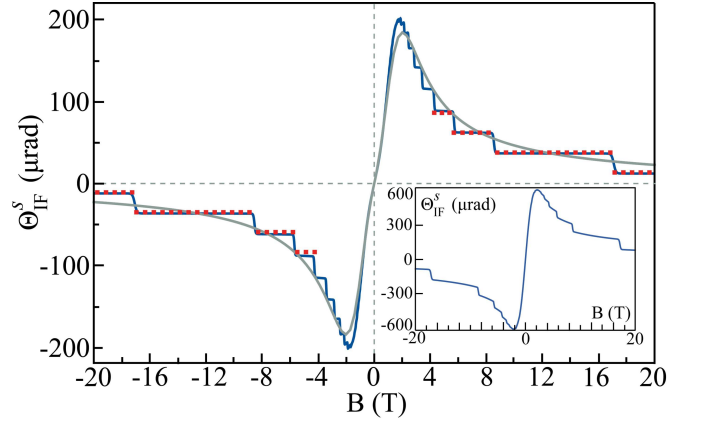}
\caption{Quantized angular IF shift for a graphene/n-doped Si system in the presence of a magnetic field. 
The red dotted curve is the quantized shift in the quantum Hall regime [Eq. (\ref{IFangularQHE})], and the solid curves correspond to the exact result for $T=4$ K (blue) and $T=300$ K (gray). The inset shows the angular IF shift for suspended graphene.
We assume an $s-$polarized incident beam with $w_0 = 1$ mm,  $\omega/2\pi = 1$ THz, and $\theta = 45^o$.
Parameters for graphene are chosen as $\mu_F = 150$ meV and $\tau = 0.184$ ps. The refractive index of undoped Si in the terahertz range is $n_{\textrm{Si}} = 3.415$, and for the doping parameters we choose 
a  carrier density of $4\times 10^{16}$ cm$^{-3}$ and a mobility of $1500$ cm$^2$/ V$\cdot$s.}
\label{Fig2}
\end{figure}

When the applied magnetic field is strong enough that quantum Hall plateaus are well-formed, for excitation frequencies $\omega$ and graphene relaxation frequency $\tau^{-1}$ much smaller than
the characteristic cyclotron frequency $\omega_c$, and $k_B T$ much smaller than the Fermi energy $\mu_F$, 
the longitudinal conductivity vanishes, 
$\sigma_{xx}^{\textrm{QHR}}= 0$, and the transverse Hall conductivity is real and quantized in multiples of the fine structure constant, 
$\sigma_{xy}^{\textrm{QHR}}  = -2(2n_c+1)\textrm{sgn}(B)e^2/2 \pi \hbar$. Here, 
$\omega_c=(\sqrt{n_c+1}-\sqrt{n_c}) \sqrt{2 e |B| v_F^2/\hbar}$ with
$n_c = \textrm{int}[\mu_F^2/ 2 \hbar e |B| v_F^2]$ the number of occupied Landau levels, and  $v_F \simeq 10^6$ m/s the Fermi velocity
\cite{Gusynin2006,Gusynin2007,Goerbig2011}. The reflection coefficients of the graphene-on-substrate system have been calculated in Ref. \cite{Tse2011, KortKamp2015}. To leading order in the fine structure constant, they are 
given as $r_{\!_{ss}} \simeq R_s $, $r_{\!_{pp}} \simeq R_p $, and 
$r_{\!_{ps}}=r_{\!_{sp}} \simeq \sigma_{xy}^{\textrm{QHR}} \sqrt{\mu_0/\varepsilon_0}  (R_p-R_s)/(\varepsilon/\varepsilon_0-1)$, 
where $R_{s,p}$ are the Fresnel reflection coefficients of the substrate and $\varepsilon_0, \mu_0$ are the vacuum permittivity and permeability. Using these expressions in Eq. (\ref{ComplexDelta}), linearly $s$- or $p$-polarized light undergoes the following angular and spatial IF shifts in the \linebreak quantum Hall regime,
\begin{eqnarray}
\label{IFangularQHE}
\Theta^{s,p}_{\rm IF}|_{_\textrm{QHR}} &=& 2 (2n_c+1)\alpha \; \textrm{sgn}(B) \Lambda^{-1}  {\rm Im}(W^{s,p}), \\
\label{IFspatialQHE}
\Delta^{s,p}_{\rm IF}|_{_\textrm{QHR}} &=& 2 (2n_c+1) \alpha \; \textrm{sgn}(B) k_0^{-1}   {\rm Re}(W^{s,p}), 
\end{eqnarray}
where $W^{s}=W^{p}|_{s\leftrightarrow p} = 2 i  \cot(\theta)  \{ |R_s|^2-|R_p|^2- 2 i {\rm Im}(R_p^* R_s) \} / |R_s|^2 (\varepsilon^*/\varepsilon_0-1)$  contain the optical properties of the substrate.
The IF shifts are quantized in integer multiples $\nu=2 (2n_c+1)=2, 6, 10, \ldots$ of the fine structure constant.
Note that the magnitude of the jumps between consecutive quantized IF shifts are independent of the optical properties of graphene and can be tuned by an appropriate choice of the substrate. 
Figure \ref{Fig2} shows the quantization of the angular IF shift in the quantum Hall regime for an $s-$polarized THz beam
impinging on a graphene-coated doped-Si substrate. 
The presence of well-defined plateaus for moderate and strong magnetic fields is clearly observed, with jumps $\simeq 24\ \mu$rad. The last plateau shows up for $B > \mu_c^2/2\hbar e v_F^2 \simeq 17$ T and gives a non-zero shift  $\Theta_{\textrm{IF}}^s\simeq 12\ \mu$rad. 
Equation (\ref{IFangularQHE}) is in excellent agreement with the result obtained using the full expressions of 
$\sigma_{xx}$ and $\sigma_{xy}$ \cite{Gusynin2006, Gusynin2007,Goerbig2011} for magnetic fields larger than $\sim 4$ T. 
Figure 3 shows the corresponding quantized spatial IF shift ${\Delta}_{\textrm{IF}}$. In this case, the magnitude of the jumps is $\simeq 182$ nm, and the last plateau gives a shift of approximately $- 91$ nm.  The full numerics agree with the prediction of Eq. (\ref{IFspatialQHE}) for magnetic fields greater than $6$ T. Gating graphene does not affect the qualitative behavior seen in the figures but simply results in a lateral distortion of \linebreak the quantized plateaus.

\begin{figure}
\centering
\hspace{-0.1cm}\includegraphics[scale=0.45]{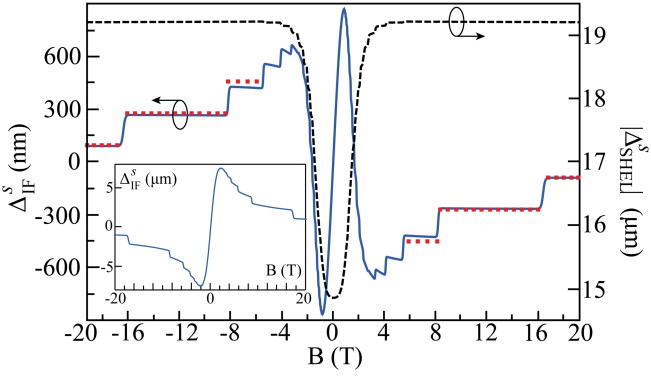}
\caption{Quantized spatial Imbert-Fedorov and photonic  spin Hall shifts as a function of the magnetic field. The dotted curve gives the spatial IF shift in the quantum Hall regime [Eq. (\ref{IFspatialQHE})] and
the solid one corresponds to the exact result. The inset shows the spatial IF shift for suspended graphene.
The dashed curve is the modulus of the relative spin Hall shift $\Delta_{\textrm{SHEL}}$. Note the different orders of magnitude of $\Delta_{\rm IF}^s $ and $|\Delta_{\textrm{SHEL}}^s|$. Temperature is set at $T=4$ K, and all other parameters are the same as in Fig. 2. }
\label{Fig3}
\end{figure}

A plateau-like behavior can still be seen in Figs. \ref{Fig2} and \ref{Fig3} for intermediate magnetic fields even though Eqs. (\ref{IFangularQHE}) and (\ref{IFspatialQHE}) no longer apply in this regime as $\omega_c$ becomes of the order of $\tau^{-1}$, meaning that disorder broadening of inter-Landau level transitions must be taken into account. To first order in $\omega\tau^{-1}/\omega_c^2$, $\sigma_{xx}$ and $\sigma_{xy}$ are non-vanishing complex quantities, and leading corrections to Eqs. (\ref{IFangularQHE}) and (\ref{IFspatialQHE}) arise from the imaginary part of the Hall conductivity
 $\sigma_{xy} \simeq \sigma_{xy}^{\textrm{QHR}}(1 + i \eta_c \omega\tau^{-1}/\omega_c^2) $, where $\eta_c = 2(\sqrt{n_c+1}-\sqrt{n_c})^2(8n_c^2+8n_c+1)/(2n_c+1)$. Disorder broadening modifies the IF shifts as 
 $\Theta_{\textrm{IF}}^{s, p} = \left.\Theta_{\textrm{IF}}^{s, p}\right|_{_\textrm{QHR}} -(\omega\tau^{-1}/\omega_c^2\Lambda) k_0 \eta_c \left.\Delta_{\textrm{IF}}^{s, p}\right|_{_\textrm{QHR}}$ and 
$\Delta_{\textrm{IF}}^{s, p} = \left.\Delta_{\textrm{IF}}^{s, p}\right|_{_\textrm{QHR}} +(\omega\tau^{-1}/\omega_c^2 k_0) \Lambda \eta_c \left.\Theta_{\textrm{IF}}^{s, p}\right|_{_\textrm{QHR}}$, resulting in a $1/|B|$ correction that can be clearly observed in Fig. 3 in the range $4 \, {\rm T} < B < 6\, {\rm T}$ as non-flat plateaus. 
For weak magnetic fields  ($B \ll \mu_F \omega / e v_F^2$), the longitudinal and Hall conductivities have a Drude form
$\sigma_{xx} =   i e^2 \mu_F / \hbar^2\pi (\omega+i\tau^{-1})$
and 
$\sigma_{xy} =  e^3 B v_F^2 / \hbar^2\pi (\omega+i\tau^{-1})^2$
(we neglect interband contributions since $\omega<2 \mu_F/\hbar$ for our parameters) \cite{Gusynin2006, Gusynin2007}.
Since $\tilde{\Delta}_{\textrm{IF}}$ is proportional to $\sigma_{xy}$ for $s-$ or $p-$ polarization, it follows that  
$\Theta_{\rm IF}^{s,p}$ grows linearly with the magnetic field at low $B$, in agreement with the numerical results shown in the figure. The shift also presents a maximum $\simeq 198\ \mu$rad around $B=1.9$ T. An increase with the magnetic field at low $B$ occurs for the spatial shift  (see Fig. 3). In contrast to ${\Theta}_{\textrm{IF}}$,  ${\Delta}_{\textrm{IF}}$ initially grows to a maximum positive value ($\simeq 860$ nm at $B=0.9$ T for the parameters in the figure), then decreases and changes sign.  

Dissipation in the substrate plays an important role in the quantized behavior of the IF shifts. For a low-loss substrate,  $W^{s,p}$ is approximately purely imaginary and hence 
$\Delta^{s,p}_{\rm IF}|_{_\textrm{QHR}} \simeq 0$ in this case. Corrections due to disorder broadening result in a non-vanishing spatial IF shift that is proportional to the angular shift, 
$\Delta_{\textrm{IF}}^{s, p} = -(\omega\tau^{-1}/\omega_c^2 k_0) \Lambda \eta_c \Theta_{\textrm{IF}}^{s, p}$. 
It is also worth mentioning that the presence of the substrate is fundamental for existence of quantized IF beam shifts. Indeed, for suspended graphene both $Y_p$ and $Y_s$ vanish since $r_{\!_{ss}} +r_{\!_{pp}}=0$ in the quantum Hall regime \cite{Tse2011}, and hence $\Theta^{s,p}_{\rm IF}|_{_\textrm{QHR}}=\Delta^{s,p}_{\rm IF}|_{_\textrm{QHR}}=0$ for suspended graphene (note that Eqs. (\ref{IFangularQHE}) and (\ref{IFspatialQHE}) are not applicable in this case). The role of the substrate is to break the symmetry between $r_{\!_{ss}}$ and  $r_{\!_{pp}}$ allowing for non-trivial quantized shifts. Still, suspended graphene does induce non-zero IF shifts in strong magnetic fields when finite-frequency and broadening corrections are taken into account. To leading order in 
$\omega/\omega_c$ and $\tau^{-1}/\omega_c$, the angular and spatial IF shifts for suspended graphene are
\begin{eqnarray}
\Theta^{s,p}_{\rm IF}|_{_\textrm{susp}} &=& \pm \sin\theta \Lambda^{-1}\textrm{sgn}(B)\zeta_c \, \tau^{-1}/\omega_c\, , \\
\Delta^{s,p}_{\rm IF}|_{_\textrm{susp}} &=& \pm \sin\theta k_0^{-1}\textrm{sgn}(B)\zeta_c \, \omega/\omega_c\, ,
\end{eqnarray}
where $\zeta_c\!\! =\!\! (4n_c+1)\sqrt{n_c+1}(\sqrt{n_c+1}\!-\!\sqrt{n_c})/(2n_c+1)$.  The shifts present non-flat plateaus, as in the graphene-on-substrate case, but they decay as $1/ \sqrt{B}$ (see insets of Figs. 2 and 3).
Finally, we briefly discuss both the role of temperature and the input frequency on the IF shifts. When $k_B T$ is no longer much smaller than the Fermi energy, the Landau level filling factors change smoothly with $B$ and the resulting $\Theta_{\rm IF}^{s,p}$ and $\Delta_{\rm IF}^{s,p}$ do not show abrupt jumps anymore but rather a continuous and mild behavior \linebreak (e.g, see Fig. 2). 
For input frequencies in the IR range or higher, quantized beam shifts do not occur as the quantum Hall condition $\omega \ll \omega_c$ does not longer hold for experimentally achievable 
magnetic fields and Fermi energies. Effects of the magnetic field on the IF shifts can still be observed in such frequency ranges, e.g. the linear increase at low $B$ and the change of sign under the reversal of the  magnetic field.  

Let us now turn our attention to the spin Hall effect of light. The reflected field in Eq. (\ref{Field}) cannot be cast as a single Gaussian beam shifted by $\Delta_{\rm IF}$ and $\Theta_{\rm IF}$. However, it can be written as a superposition of displaced Gaussians by introducing the left $(+)$ and right $(-)$ circularly polarized 
basis $\hat{{\bf e}}_{\pm} = [\hat{{\bf x}}_r\pm i(\hat{{\bf y}}_r-Cy_r\hat{{\bf z}}_r)]/\sqrt{2}$:
\begin{eqnarray}
\!\!\!\!\!{\bf E}_r\! &=&\!  \sqrt{2}  (f_p r_{\!_{pp}} + f_s  r_{\!_{ps}} )  \cr
\!\!&\times& \!\! \Big\{\!\!\left[(1\!-\!im_R)A(y_r\!-\!\tilde{\delta}_2,z_r) + m_IA(y_r\!-\!\tilde{\delta}_1,z_r) \right] \hat{{\bf e}}_{+}  \cr
\!\!&+&\! \left[(1\!+\!im_R)A(y_r\!-\!\tilde{\delta}_1,z_r) - m_IA(y_r\!-\!\tilde{\delta}_2,z_r) \right] \hat{{\bf e}}_{-}\!  \Big\} ,
\label{FieldSHEL}
\end{eqnarray}
where $m_R = \textrm{Re}(m)$, $m_I = \textrm{Im}(m)$. The complex displacements $\tilde{\delta}_l $ are given by
$\tilde{\delta}_l = \tilde{\Delta}_{\textrm{IF}} + (-1)^l \tilde{\Delta}_{\textrm{SHEL}}$ $(l = 1, 2)$, where 
$\tilde{\Delta}_{\textrm{SHEL}}$ is obtained from Eq. (\ref{ComplexDelta}) replacing $Y_p$ by $ imY_p$ and $Y_s$ by $-iY_s/m^{*}$.  Note that each polarized state of the reflected field is a superposition of two Gaussians centered at distinct positions.
The SHEL spatial shifts for each polarization state are computed from the intensity distribution centroid 
for the $\hat{{\bf e}}_{+}$ and $\hat{{\bf e}}_{-}$ components of the field, and are given by $\delta^{\pm}={\rm Re} [\tilde{\delta}^{\pm}]$, where
\begin{equation}
\tilde{\delta}^{+} =
\frac{(1+m_R^2) \tilde{\delta}_{2} + m_I^2 \tilde{\delta}_{1} +  m_I (1 - i m_R) 
(\tilde{\delta}_{1}^* + \tilde{\delta}_{2})}{1+ |m|^2},
\end{equation}
and $\tilde{\delta}^{-}$ is obtained by swapping $1 \leftrightarrow 2$ and replacing $m$ by $-m$ in the above equation.
Note that the shifts for each polarization are given as a weighted average of $\tilde{\delta}_1$ and $\tilde{\delta}_2$ plus an overlap term.
In the low-dissipation limit ($m_I \ll m_R$), the reflected polarized components in Eq. (\ref{FieldSHEL}) reduce to single Gaussians, and right- and left-polarized photons are respectively shifted by $\tilde{\delta}_1$ and $\tilde{\delta}_2$, as in \cite{Bliokh2006}. It is useful to express the SHEL shifts referred to the spatial IF shift of the whole reflected beam, $\delta^{\pm} = \Delta_{\textrm{IF}} \pm \Delta_{\textrm{SHEL}}$, where the SHEL relative shift is
$\Delta_{\textrm{SHEL}} = \textrm{Re}[ \tilde{\Delta}_{\textrm{SHEL}}(1+m^{*2})+ 2m_I \tilde{\Delta}_{\textrm{IF}}]/ (1+|m|^2)$.
In contrast to the spatial IF shift $\Delta_{\textrm{IF}}$, the relative SHEL shift is an even function of the magnetic field.

In Figure 3 we plot the SHEL relative shift as a function of magnetic field for our graphene-substrate system.
In the quantum Hall regime, the SHEL shifts for incident $s-$ or $p-$ polarized light are the sum of the quantized spatial IF shift [eq. (\ref{IFspatialQHE})] and the relative SHEL shift $\Delta_{\textrm{SHEL}}^{s,p} |_{\textrm{QHR}} =-\cot(\theta) \textrm{Re}[(R_p+R_s)/R_{s,p}]/k_0$. Note that this second term depends only on the substrate optical properties, is much larger than $\Delta_{\textrm{IF}}|_{\textrm{QHR}}$
as it is independent of the fine structure constant, and is actually the SHEL shift for an air-substrate interface in the absence of the graphene coating \cite{Bliokh2006}. For intermediate magnetic fields disorder broadening in graphene results in non-flat plateaus as in the case for the spatial IF shift, and at low fields $|\Delta_{\textrm{SHEL}}^{s,p}|$ present a minimum at zero field, precisely \linebreak where $\Delta_{\textrm{IF}}$
vanishes.
\begin{figure}
\centering
\includegraphics[scale=0.47]{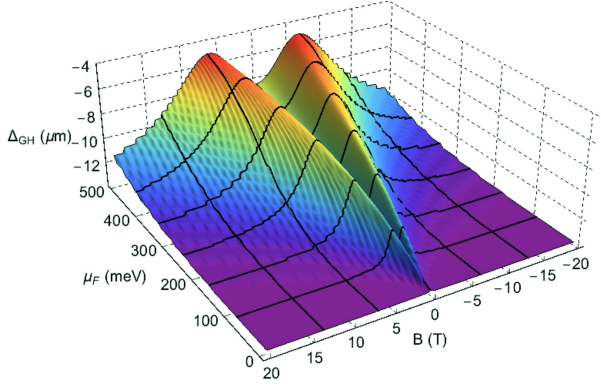}
\caption{Spatial Goos-H\"anchen shift for the graphene-coated substrate as a function of the magnetic field and Fermi energy of graphene. Temperature is set at $T=4$ K, and all other parameters are the same as in Fig. 2.}
\label{Fig4}
\end{figure}

Finally, we compute the spatial and angular GH shifts. To this end, we assume that the impinging Gaussian beam is confined within the plane of incidence. Performing analogous calculations as for the IF shift, one can show that the GH ones are 
$\Delta_{\textrm{GH}} = \textrm{Re}[\tilde{\Delta}_{\textrm{GH}}]$ and $\Theta_{\textrm{GH}} = (k_0/\Lambda)\textrm{Im}[\tilde{\Delta}_{\textrm{GH}}]$. Here,  $\tilde{\Delta}_{\textrm{GH}}$ is given by equation (\ref{ComplexDelta}) with $Y_{s, p}$ replaced by $ X_{s, p}$ and $X_{p} = {X_s}|_{s\leftrightarrow p} = -i(f_p \partial_{\theta} r_{\!_{pp}} + f_s \partial_{\theta} r_{\!_{ps}})/k_0r_{\!_{pp}}$. In the quantum Hall regime the co-polarized terms in $X_s$ and $X_p$ give contributions to the GH shifts that are independent of the optical properties of graphene
and correspond to the usual shifts $\Delta_{\textrm{GH, sub}}^{s,p}=k_0^{-1} {\rm Im}[\partial_{\theta} \log R_{s,p}]$ and
$\Theta_{\textrm{GH, sub}}^{s,p}=- \Lambda^{-1} {\rm Re}[\partial_{\theta} \log R_{s,p}]$ for uncoated isotropic substrates \cite{Bliokh2013}.
The cross-polarized reflection coefficients in $X_s$ and $X_p$ bring about the influence of the electronic quantum Hall effect of graphene on the GH shifts. The full GH shifts in this regime are then given as
$\Delta_{\textrm{GH}}^{s,p} |_{\textrm{QHR}} =\Delta_{\textrm{GH,sub}}^{s,p} - 4 (2n_c+1)^2\alpha^2k_0^{-1}\textrm{Re}({\cal{K}}^{s,p})$ and $\Theta^{s,p}_{\textrm{GH}} |_{\textrm{QHR}} = \Theta_{\textrm{GH, sub}}^{s,p} -4(2n_c+1)^2\alpha^2\Lambda^{-1}\textrm{Im}({\cal{K}}^{s,p})$ where ${\cal{K}}^{s} = {\cal{K}}^{p}|_{s\leftrightarrow p}\! =\! 4i(R_p^*-R_s^*)(\partial_{\theta} R_p-\partial_{\theta}R_s)/|R_s|^2|\varepsilon/\varepsilon_0-1|^2 $. Therefore, the GH shifts are quantized functions of the magnetic field. In contrast to the IF quantized plateaus, the GH ones are much weaker ($\propto \alpha^2$). Both the spatial and angular GH shifts are even functions of the magnetic field. Figure \ref{Fig4} shows the spatial GH shift as a function of the magnetic field and graphene's Fermi energy. Plateau-like behavior can be observed both tuning $B$ or $\mu_F$. The Drude-like response of graphene at low $B$ results in a quadratic dependency of the GH shifts on the magnetic field.

Previous experimental demonstrations of beam shifts in the optical range have been accomplished for ratios $\Delta_{\rm IF}/w_0$ as small as $10^{-6}$ to $10^{-4}$ \cite{Hosten2008,Aiello2010}. In our case, the minimal ratio $\Delta_{\rm IF}/w_0$  to resolve quantized beam shifts should be on the order of $2 \times 10^{-4}$  for the parameters used in Fig. 3. 
Given the recent advances in THz lasers, detectors, and optical elements \cite{Tonouchi2007},
the demonstration of quantized shifts is within experimental reach. 

In conclusion, our studies reveal a plethora of novel magneto-optical effects that ultimately originate from the chiral properties of electrons in Landau levels. We predict the quantization of beams shifts in graphene-coated materials. The resulting discrete shifts in the quantum Hall regime allow for a precise control of the spatial and polarization distributions of the reflected and transmitted beams. We envision that the effects predicted in this work could be enhanced by using graphene metasurfaces thanks to their ability to tailor both the magneto-optical response of graphene and the spin-orbit coupling of photons. 

We would like acknowledge P. W. Milonni and H.-T. Chen for discussions and the LANL LDRD program for financial support.
\end{document}